\documentclass[
 amsmath,amssymb,
 aps,
]{revtex4-2}

\usepackage{graphicx}
\usepackage{dcolumn}
\usepackage{bm}
\usepackage{amsmath}
\usepackage{slashed}
\usepackage{revsymb}

\makeatletter

\newcommand\makebig[2]{%
  \@xp\newcommand\@xp*\csname#1\endcsname{\bBigg@{#2}}%
  \@xp\newcommand\@xp*\csname#1l\endcsname{\@xp\mathopen\csname#1\endcsname}%
  \@xp\newcommand\@xp*\csname#1r\endcsname{\@xp\mathclose\csname#1\endcsname}%
}
\makeatother

\makebig{biggg} {4.0}
\makebig{Biggg} {4.5}
\makebig{bigggg}{5.0}
\makebig{Bigggg}{5.5}

\begin{document}

\title{Electron Wave-Spin Qubit}
\author{Ju Gao}
\email{jugao@illinois.edu}
\author{Fang Shen}%
\affiliation{University of Illinois, Department of Electrical and Computer Engineering, Urbana, 61801, USA}
\date{\today}

\begin{abstract}
As a continuation of our earlier investigations into electron wave--spin~\citep{GaoJOPCO22,EntropyEvaSpin2024}, 
we analyze the electron spin and its qubit in a cavity by treating the electron as a physical wave obeying the Dirac equation. 
In this view, a qubit is a current--density configuration whose orientation is fixed by the relative phase, 
rather than a particle carrying simultaneous ``up'' and ``down'' spin states with assigned probabilities. 
The resulting magnetic--moment density, derived from the current, displays a richer vector distribution and topology 
than the fixed axial dipole weighted by probability density in the conventional wave--particle model. 
Both frameworks yield the same total moment of one Bohr magneton and are indistinguishable in uniform external fields, 
yet their ontological differences predict distinct couplings to structured fields and spin--spin interactions. 
These contrasts motivate further exploration of dynamical consequences within the wave--entity framework, 
including Aharonov--Bohm--like responses that provide testable alternatives to conventional wave--particle duality.
\end{abstract}


\maketitle

\section{\label{sec:Intro}Introduction}
The electron, a cornerstone of modern physics and technology, remains enigmatic. 
Conventionally modeled as a point-like particle of charge $q_{\!e}=-e=-1.602\times10^{-19}~\mathrm{C}$, 
its electromagnetic behavior is described by the Lorentz--covariant four--current 
$j^\mu=(c\,q(\bm r,t),\,\bm j(\bm r,t))$, where $q(\bm r,t)$ and $\bm j(\bm r,t)$ denote 
charge and current densities~\cite{JDJackson1999}. 
This four--current governs both the generation of and interaction with the electromagnetic field 
$A^\mu=(\phi/c,\,\bm A)$, with scalar and vector potentials $\phi$ and $\bm A$:
\begin{equation}
\square A^\mu = \mu_{0} j^\mu, 
\qquad
\mathcal{L}_{\text{int}} = - j^{\mu}A_{\mu}=-\,q\,\phi + \bm{j}\!\cdot\!\bm{A},
\label{GenerationAndInteraction}
\end{equation}
where $\square$ is the d'Alembertian and we adopt the Minkowski metric $(+,-,-,-)$, consistent with $-j^{\mu}A_{\mu} = -\,q\,\phi + \bm{j}\!\cdot\!\bm{A}$.

Electrons also exhibit unmistakably wave--like behavior, from double--slit interference~\cite{jonsson1961elektroneninterferenzen,jonsson1961electron,tonomura1989demonstration} to diffraction~\cite{davisson1927diffraction,thomson1927diffraction}. Recent experimental progress highlights that spin qubits in silicon and related platforms can now achieve high fidelity even above $1~\mathrm{K}$~\cite{Huang2024Nature}, scalable integration~\cite{Hu2025iComputing,Meunier2025EPJA}, and advanced control via engineered magnetic textures~\cite{Cai2025PRApplied}.

A comprehensive quantum description is therefore required, in which the four--current is expressed through the wavefunction $\Psi(\bm{r},t)$:
\begin{equation}\label{fourcurrent}
q(\bm{r},t)=-e\,\Psi^{\dagger}(\bm{r},t)\Psi(\bm{r},t),
\qquad
\bm{j}(\bm{r},t)=-e c\,\Psi^{\dagger}(\bm{r},t)\bm{\alpha}\Psi(\bm{r},t).
\end{equation}
The wavefunction evolves according to the Dirac equation~\cite{Dirac28,dirac1930principles}:
\begin{equation}\label{Dirac}
i\hbar\,\partial_{t}\Psi(\bm{r},t)
=\Big[-i\hbar c\,\bm{\alpha}\!\cdot\!\nabla + \gamma^{0} m_{e} c^{2}\Big]\Psi(\bm{r},t),
\end{equation}
with electron mass $m_{e}$, speed of light $c$, and reduced Planck constant $\hbar$. The Dirac formalism unifies charge, current, and spin within a single Lorentz--covariant wave equation.

For an energy eigenstate $\Psi(\bm{r},t)=e^{-i\mathcal{E}t/\hbar}\,\Psi(\bm{r})$, the current density can be written~\cite{Ohanian86,GaoJOPCO22}:
\begin{equation}\label{current}
\bm{j}(\bm{r})=-\frac{e c^{2}}{\mathcal{E}}
\!\left[
\nabla\!\times\!\Big(\Psi^{\dagger}(\bm{r})\,\tfrac{\hbar}{2}\bm{\Sigma}\,\Psi(\bm{r})\Big)
+i\tfrac{\hbar}{2}\Big((\nabla\Psi^{\dagger})\Psi-\Psi^{\dagger}\nabla\Psi\Big)
\right],
\end{equation}
where $\bm{\Sigma}=\tfrac{1}{2i}\,\bm{\alpha}\!\times\!\bm{\alpha}$ is the spin operator. The first term represents a circulating spin--associated current; the second, translational motion. Thus the electron wave inherently carries spin, even in the absence of external fields. The observed spin value $\hbar/2$ arises after integrating spatial interactions with magnetic fields~\cite{GaoJOPCO22}, indicating that spin is not a localized attribute but a structured, intrinsic feature of the electron wave---here termed \emph{wave--spin} to distinguish it from the conventional particle--spin picture.

From this standpoint, the electron wave itself emerges as a natural candidate for the fundamental entity,
carrying charge, current, and spin in a Lorentz--covariant and self--consistent manner. 
This perspective avoids attributing a hidden particle structure to the electron and avoids reducing the wave to a purely statistical abstraction.

Crucially, the \emph{wave--entity} (a real, spatially extended object) is not identical to the abstract Hilbert--space wavefunction. 
The wavefunction remains a mathematical device that evolves deterministically under the Dirac equation and computes observables; 
the wave--entity is the physical carrier whose charge and current densities are expressed through the wavefunction. 
What is usually called ``collapse of the wavefunction'' is reinterpreted as a transition of the wave--entity 
from one physical configuration to another, not the disappearance of a real object.

Within this framework, no localized ``particle electron'' with definite size or shape exists. 
Apparent particle--like behaviors---such as cathode--ray impacts or electron--beam lithography---arise 
from the small collision cross--section of a real wave interacting with matter, 
in close analogy with electromagnetic scattering.

This wave--based ontology depicts the electron as a deterministic physical wave evolving under relativistic equations, 
positioning the wave--rather than the particle--as the basic entity for quantum devices. 
Using this framework, we examine electron spin and qubits by solving the Dirac equation in a cylindrical cavity 
and representing spin--qubit states through current--density configurations. 
We then analyze their interaction with external fields by calculating the magnetic--moment density 
and comparing it with the corresponding quantity in the conventional wave--particle view. These distinctions connect naturally to Aharonov--Bohm--like responses, where modern treatments have explored complex vector potentials~\cite{Paiva2023NJP} and spin--dependent analogues of the AB effect~\cite{Chen2023SpinAB}. Together, these results provide a concrete basis for experimental verification and 
naturally motivate Aharonov--Bohm--like tests of structured--field couplings, which we pursue in companion work. 

\section{\label{sec:ElectronCavity}Electron wave--spin in a cavity}
We derive analytical wavefunctions and current densities of an electron confined in a finite cylindrical quantum dot, modeled as a three--dimensional cavity with partial radial confinement and complete axial confinement. The confined Dirac electron exhibits a toroidal wave--spin topology, which differs from the conventional particle--based interpretation of spin and demonstrates spin as an extended spatial structure rather than an intrinsic point--like attribute.

We begin with the Dirac equation in a cylindrically symmetric potential:
\begin{equation} \label{Diracwithpotential}
i\hbar \frac{\partial}{\partial t}\Psi (\bm{r},t)
=\left[ -i\hbar c \,\bm{\alpha} \cdot \bm{\nabla }+\gamma ^0 m_{e} c^2+U(\bm{r})\right]\Psi(\bm{r},t),
\end{equation}
where $U(\bm{r})$ models the confinement~\cite{loss1998quantum,hanson2007spins}:
\begin{equation}\label{potential}
U(\bm{r})=\begin{cases}
0, & 0<\rho<R,\; -d<z<d \quad \text{(Region I)},\\
U, & \rho>R,\; -d<z<d \quad \text{(Region II)},\\
\infty, & z<-d \text{ or } z>d \quad \text{(Region III)}.
\end{cases}
\end{equation}

The Dirac momentum operator in cylindrical coordinates $(\rho,\phi,z)$ is
\begin{equation}\label{momentum}
\bm{\alpha} \cdot \bm{\nabla}
=\alpha _{\rho }\frac{\partial }{\partial \rho }
+\alpha _{\phi }\frac{1}{\rho }\frac{\partial }{\partial \phi }
+\alpha _z\frac{\partial }{\partial z},
\end{equation}
with $\alpha_\rho=\alpha_x\cos\phi+\alpha_y\sin\phi$ and 
$\alpha_\phi=-\alpha_x\sin\phi+\alpha_y\cos\phi$ so that 
$\rho^{-1}\partial_\phi$ acts only on the spinor components while the $\phi$--dependence of the basis vectors is encoded in $\alpha_\rho$ and $\alpha_\phi$. 
Together with $\alpha_z$, they satisfy
\begin{equation}\label{alpha}
\alpha_{\rho }^2=\alpha_{\phi }^2=\alpha_{z}^2=\mathbb{1},\quad
[\alpha _{\rho },\alpha_{\phi }]=2i \Sigma_z,\quad
\{ \alpha_{\rho },\alpha_{\phi }\}=0.
\end{equation}

The electron is fully confined along $z$ but only partially confined radially, giving rise to an evanescent wave outside the cavity~\cite{EntropyEvaSpin2024}. While Schrödinger solutions for related geometries are known for planar systems~\cite{lobanova2004cylindrical}, we solve the full Dirac equation to expose the relativistic wave--spin structure.

Assuming a stationary state $\Psi(\bm{r},t)=e^{-i\mathcal{E}t/\hbar}\,\psi(\rho,\phi,z)$ with eigenenergy $\mathcal{E}$, we solve for $\psi$ in Regions I and II (vanishing in Region III). As established in Ref.~\cite{EntropyEvaSpin2024}, the interior solution remains accurate despite evanescent components at the boundary.

By separation of variables, we obtain eigenfunctions for spin--up and spin--down, labeled by $(n,l,m)$. For brevity we display the general forms and then specialize to the ground state. The spin--up state $\psi_{nlm\uparrow}$ is
\begin{widetext}
\begin{equation}\label{psinlmup}
\psi_{nlm\uparrow}(\rho,\phi,z)=
\begin{cases}
\displaystyle
\phantom{\kappa_{nlm}}N \begin{pmatrix}
J_l(\zeta_{nlm}\rho)\,e^{il\phi}\cos(k_m z) \\[1ex]
0 \\[1ex]
i\,\eta_{\mathrm{I}}\,k_m\,J_l(\zeta_{nlm}\rho)\,e^{il\phi}\sin(k_m z) \\[1ex]
-i\,\eta_{\mathrm{I}}\!\left[\tfrac{\zeta_{nlm}}{2}\!\Big(J_{l-1}(\zeta_{nlm}\rho)-J_{l+1}(\zeta_{nlm}\rho)\Big)-\tfrac{l}{\rho}J_l(\zeta_{nlm}\rho)\right] e^{i\phi}e^{il\phi}\cos(k_m z)
\end{pmatrix}, & \text{Region I},\\[2.0ex]
\displaystyle
\kappa_{nlm} N \begin{pmatrix}
K_l(\xi_{nlm}\rho)\,e^{il\phi}\cos(k_m z) \\[1ex]
0 \\[1ex]
i\,\eta_{\mathrm{II}}\,k_m\,K_l(\xi_{nlm}\rho)\,e^{il\phi}\sin(k_m z) \\[1ex]
-i\,\eta_{\mathrm{II}}\!\left[\tfrac{\xi_{nlm}}{2}\!\Big(-K_{l-1}(\xi_{nlm}\rho)-K_{l+1}(\xi_{nlm}\rho)\Big)-\tfrac{l}{\rho}K_l(\xi_{nlm}\rho)\right] e^{i\phi}e^{il\phi}\cos(k_m z)
\end{pmatrix}, & \text{Region II},
\end{cases}
\end{equation}
\end{widetext}

The spin--down state $\psi_{nlm\downarrow}$ is

\begin{widetext}
\begin{equation}\label{psinlmdown}
\psi_{nlm\downarrow}(\rho,\phi,z)=
\begin{cases}
\displaystyle
\phantom{\kappa_{nlm}}N \begin{pmatrix}
0 \\[1ex]
J_l(\zeta_{nlm}\rho)\,e^{il\phi}\cos(k_m z) \\[1ex]
-i\,\eta_{\mathrm{I}}\!\left[\tfrac{\zeta_{nlm}}{2}\!\Big(J_{l-1}(\zeta_{nlm}\rho)-J_{l+1}(\zeta_{nlm}\rho)\Big)+\tfrac{l}{\rho}J_l(\zeta_{nlm}\rho)\right] e^{-i\phi}e^{il\phi}\cos(k_m z) \\[1ex]
-i\,\eta_{\mathrm{I}}\,k_m\,J_l(\zeta_{nlm}\rho)\,e^{il\phi}\sin(k_m z)
\end{pmatrix}, \; \text{Region I},\\[2.0ex]
\displaystyle
\kappa_{nlm} N \begin{pmatrix}
0 \\[1ex]
K_l(\xi_{nlm}\rho)\,e^{il\phi}\cos(k_m z) \\[1ex]
-i\,\eta_{\mathrm{II}}\!\left[\tfrac{\xi_{nlm}}{2}\!\Big(-K_{l-1}(\xi_{nlm}\rho)-K_{l+1}(\xi_{nlm}\rho)\Big)+\tfrac{l}{\rho}K_l(\xi_{nlm}\rho)\right] e^{-i\phi}e^{il\phi}\cos(k_m z) \\[1ex]
-i\,\eta_{\mathrm{II}}\,k_m\,K_l(\xi_{nlm}\rho)\,e^{il\phi}\sin(k_m z)
\end{pmatrix}, \; \text{Region II}.
\end{cases}
\end{equation}
\end{widetext}

Here $e^{il\phi}$ encodes the azimuthal quantum number $l=0,1,2,\dots$, and $k_m=\frac{m\pi}{2d}$ with $m=1,3,5,\dots$. The radial functions are Bessel $J_l$ and modified Bessel $K_l$.
The radial wave numbers relate to $\mathcal{E}_{nlm}$ by
\begin{equation}\label{zetaxi}
\zeta_{nlm}^2=\frac{\mathcal{E}_{nlm}^2-m_{e}^2 c^4-\hbar ^2 c^2 k_m^2}{\hbar ^2 c^2},
\qquad
\xi_{nlm}^2=\frac{(\mathcal{E}_{nlm}-U)^2-m_{e}^2 c^4-\hbar ^2 c^2 k_m^2}{\hbar ^2 c^2 }.
\end{equation}
The geometric factors are
\begin{equation}\label{eta}
\eta _{\mathrm{I}}=\frac{\hbar c}{\mathcal{E}_{nlm}+m_{e} c^2},
\qquad
\eta _{\mathrm{II}}=\frac{\hbar c}{\mathcal{E}_{nlm}-U+m_{e} c^2},
\end{equation}
which converge in the nonrelativistic limit to
\begin{equation}\label{eta2}
\eta _{\mathrm{I}}\approx \eta _{\mathrm{II}}\approx \eta=\frac{\hbar}{2m_{e}c}.
\end{equation}

At $\rho=R$, continuity yields
\begin{equation}\label{boundarynlm}
J_l(\zeta_{nlm} R ) = \kappa_{nlm} K_l(\xi_{nlm} R),\qquad
\eta_{\mathrm{I}}\!\left[\zeta_{nlm} J_{l}'(\zeta_{nlm} R)-\frac{l}{R}J_l(\zeta_{nlm}R)\right]
=\kappa_{nlm}\eta_{\mathrm{II}}\!\left[\xi_{nlm} K_{l}'(\xi_{nlm} R)-\frac{l}{R}K_l(\xi_{nlm}R)\right],
\end{equation}
which together determine $\zeta_{nlm}$ (and hence $\xi_{nlm}$, $\kappa_{nlm}$, and $\mathcal{E}_{nlm}$).

For the ground state $(n,l,m)=(1,0,1)$ we write $k_1\equiv k=\frac{\pi}{2d}$, $\zeta_{101}\equiv\zeta$, $\xi_{101}\equiv\xi$, and $\kappa_{101}\equiv\kappa$. The spin--up and spin--down wavefunctions are
\begin{widetext}
\begin{equation}\label{psi101up}
\psi_{\uparrow}(\rho,\phi,z)=
\begin{cases}
\displaystyle
\phantom{\kappa}N\begin{pmatrix}
J_0(\zeta \rho )\cos(kz) \\
0 \\
i\,\eta\,k\,J_0(\zeta \rho )\sin(k z) \\
i\,\eta\,\zeta\,J_1(\zeta \rho ) e^{i \phi}\cos(kz)
\end{pmatrix}, & \text{Region I},\\[2.0ex]
\displaystyle
\kappa N\begin{pmatrix}
K_0(\xi \rho)\cos(kz) \\
0 \\
i\,\eta\,k\,K_0(\xi \rho )\sin(kz) \\
i\,\eta\,\xi\,K_1(\xi \rho)e^{i \phi}\cos(kz)
\end{pmatrix}, & \text{Region II},
\end{cases}
\end{equation}
\end{widetext}

\begin{widetext}
\begin{equation}\label{psi101down}
\psi_{\downarrow}(\rho,\phi,z)=
\begin{cases}
\displaystyle
\phantom{\kappa}N\begin{pmatrix}
0 \\
J_0(\zeta \rho )\cos(kz) \\
i\,\eta\,\zeta\,J_1(\zeta \rho ) e^{-i \phi}\cos(kz) \\
-\,i\,\eta\,k\,J_0(\zeta \rho )\sin(k z)
\end{pmatrix}, & \text{Region I},\\[2.0ex]
\displaystyle
\kappa N\begin{pmatrix}
0 \\
K_0(\xi \rho)\cos(kz) \\
i\,\eta\,\xi\,K_1(\xi \rho)e^{-i \phi}\cos(kz) \\
-\,i\,\eta\,k\,K_0(\xi \rho )\sin(kz)
\end{pmatrix}, & \text{Region II}.
\end{cases}
\end{equation}
\end{widetext}

These states are degenerate and can be superposed without energy splitting, enabling qubit constructions without beating.

The ground--state boundary conditions reduce to
\begin{equation}\label{boundary101}
J_0(\zeta R)=\kappa\,K_0(\xi R),
\qquad
\zeta\,J_1(\zeta R)=\kappa\,\xi\,K_1(\xi R).
\end{equation}
Normalization ($\int\Psi^\dagger\Psi\,d^3r=1$) gives
\begin{equation}\label{N2}
N^2=\frac{1}{\pi R^2 d \,\big[J_1(\zeta R)^2 + \kappa^2 K_1(\xi R)^2\big]} .
\end{equation}

The current density for the spin--up ground state is
\begin{equation}\label{current101}
j_\phi(\rho,z)=
\begin{cases}
-\,2 N^2 e c\,\eta\,\zeta \, J_0(\zeta \rho)\, J_1(\zeta \rho)\cos^2(k z), & \text{Region I},\\
-\,2 \kappa^2 N^2 e c\,\eta\,\xi \, K_0(\xi \rho)\, K_1(\xi \rho)\cos^2(k z), & \text{Region II},
\end{cases}
\qquad
j_\rho(\rho,z)=0,\;\; j_z(\rho,z)=0.
\end{equation}
The azimuthal current highlights the wave--spin character of the electron, forming a toroidal topology even for $l=0$. Moreover, the current extends beyond the cavity boundary, demonstrating an evanescent wave--spin.

\begin{figure}
\includegraphics[width=0.38\textwidth]{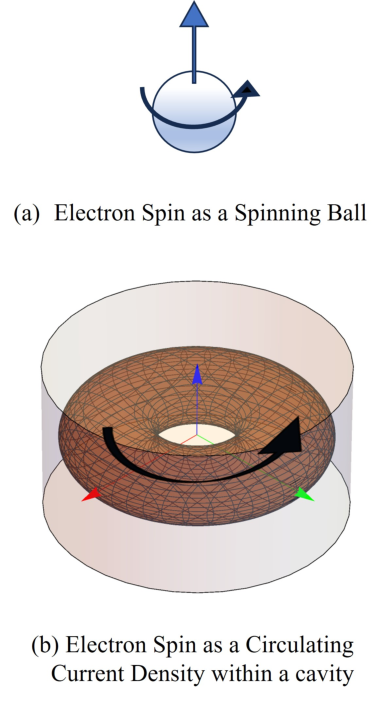}%
\caption{\label{fig:fig1} 
(a) Conventional particle--spin model, depicting the electron as a rotating corpuscular ball. 
(b) Toroidal contour of the current density plotted at two--thirds of its peak value. 
The electron is confined within a cylindrical cavity of radius $R = 8~\text{nm}$, height $d = 4~\text{nm}$, and potential energy $U = 10~\text{meV}$. 
The spin--up and spin--down components are degenerate in the ground state $(n l m = 101)$, with eigenenergy $\mathcal{E}_{101} - m_{e}c^{2} = 8.06~\text{meV}$.}
\end{figure}

For $R = 8~\text{nm}$, $d = 4~\text{nm}$, and $U = 10~\text{meV}$, solving Eq.~\ref{boundary101} yields the ground--state energy $\mathcal{E}_{101} - m_{e}c^{2} = 8.06~\text{meV}$. These parameters fix $\zeta$, $\xi$, $\kappa$, $N$ via Eqs.~\ref{zetaxi}, \ref{eta}, and \ref{N2}, enabling three--dimensional contour visualization of the current density (Fig.~\ref{fig:fig1}). For contrast, Fig.~\ref{fig:fig1}(a) depicts the particle--spin model.

Revealing the toroidal topology underscores that the electron wave is a spatially extended physical entity; reducing it to a point discards physically meaningful structure. Such topology encodes persistent features that may be comparatively robust to perturbations, potentially informing studies of long--lived coherence~\cite{Lambert2013QuantumBiology} and the design of stable, fault--tolerant spin qubits. This perspective also aligns with recent comprehensive analyses of decoherence mechanisms in solid--state spin qubits~\cite{Onizhuk2025RMP}, underscoring the importance of physically grounded models for understanding robustness.

\section{\label{sec:CavitySpinCylinder}Electron wave--spin qubit}
We now examine the electron spin qubit within the wave--entity framework, which is both a basic physics problem and a cornerstone of quantum computing.

The spin--qubit state is the superposition
\begin{equation}\label{eq:Qubit}
\psi=\cos\!\frac{\Theta}{2}\,\psi_{\uparrow}
      +\sin\!\frac{\Theta}{2}\,e^{i\Phi}\,\psi_{\downarrow},
\end{equation}
where $0\le \Theta\le \pi$ and $0\le \Phi\le 2\pi$ locate the state on the Bloch sphere~\cite{bloch1946nuclear}, and logical operations act as sphere rotations~\cite{nielsen2010quantum}. The parameter $\Theta$ sets the amplitudes, and the relative phase $\Phi$ governs interference between basis states. The physical significance of phase control in qubits has been further underscored in recent demonstrations of coherent control and coupling in multielectron spin states~\cite{Song2024npjQI} and in baseband-driven qubit architectures~\cite{Unseld2025NatComm}.

The centrality of $\Phi$ in quantum information is well established: in the quantum Fourier transform~\cite{cleve1998quantum,coppersmith2002approximate} phases encode periodicity exploited by Shor’s algorithm~\cite{shor1999polynomial}; in Grover’s search~\cite{grover1997quantum}, phase inversion drives interference; and many algorithms hinge on phase control~\cite{nielsen2002quantum,deutsch1992rapid}. Within the wave--entity picture this abstract role becomes concrete: the phase~$\Phi$ enters directly into the current density, and thus into the field interaction, 
tying qubit phases to measurable real--space flow.

Combining Eqs.~\ref{psi101up}, \ref{psi101down}, and \ref{eq:Qubit} yields (Region I/II forms shown compactly)
\begin{widetext}
\begin{eqnarray}\label{eq:Qubit_current}
j_\rho&=&
\begin{cases}
-\,N^2 e c \,\eta \,\sin\Theta \,\sin(\Phi-\phi)\,k\,J_0(\zeta \rho)^2\,\sin(2kz), & \text{Region I},\\
-\,\kappa^2 N^2 e c \,\eta \,\sin\Theta \,\sin(\Phi-\phi)\,k\,K_0(\xi \rho)^2\,\sin(2kz), & \text{Region II},
\end{cases}\nonumber\\[0.4ex]
j_\phi&=&
\begin{cases}
-\,2N^2 e c \,\eta \,\cos\Theta \,\zeta\,J_0(\zeta \rho)J_1(\zeta \rho)\cos^2(kz)
+N^2 e c \,\eta \,\sin\Theta \,\cos(\Phi-\phi)\,k\,J_0(\zeta \rho)^2\,\sin(2kz), & \text{Region I},\\
-\,2\kappa^2 N^2 e c \,\eta \,\cos\Theta \,\xi\,K_0(\xi \rho)K_1(\xi \rho)\cos^2(kz)
+\kappa^2 N^2 e c \,\eta \,\sin\Theta \,\cos(\Phi-\phi)\,k\,K_0(\xi \rho)^2\,\sin(2kz), & \text{Region II},
\end{cases}\nonumber\\[0.4ex]
j_z&=&
\begin{cases}
2N^2 e c \,\eta \,\sin\Theta \,\sin(\Phi-\phi)\,\zeta\,J_0(\zeta \rho)J_1(\zeta \rho)\cos^2(kz), & \text{Region I},\\
2\kappa^2 N^2 e c \,\eta \,\sin\Theta \,\sin(\Phi-\phi)\,\xi\,K_0(\xi \rho)K_1(\xi \rho)\cos^2(kz), & \text{Region II}.
\end{cases}
\end{eqnarray}
\end{widetext}
For the equal superposition $\Theta=\pi/2$,
\begin{widetext}
\begin{eqnarray}\label{eq:Qubit_current_equator}
j_\rho&=&
\begin{cases}
-\,N^2 e c \,\eta \,\sin(\Phi-\phi)\,k\,J_0(\zeta \rho)^2\,\sin(2kz), & \text{Region I},\\
-\,\kappa^2 N^2 e c \,\eta \,\sin(\Phi-\phi)\,k\,K_0(\xi \rho)^2\,\sin(2kz), & \text{Region II},
\end{cases}\nonumber\\[0.4ex]
j_\phi&=&
\begin{cases}
N^2 e c \,\eta \,\cos(\Phi-\phi)\,k\,J_0(\zeta \rho)^2\,\sin(2kz), & \text{Region I},\\
\kappa^2 N^2 e c \,\eta \,\cos(\Phi-\phi)\,k\,K_0(\xi \rho)^2\,\sin(2kz), & \text{Region II},
\end{cases}\nonumber\\[0.4ex]
j_z&=&
\begin{cases}
2N^2 e c \,\eta \,\sin(\Phi-\phi)\,\zeta\,J_0(\zeta \rho)J_1(\zeta \rho)\cos^2(kz), & \text{Region I},\\
2\kappa^2 N^2 e c \,\eta \,\sin(\Phi-\phi)\,\xi\,K_0(\xi \rho)K_1(\xi \rho)\cos^2(kz), & \text{Region II}.
\end{cases}
\end{eqnarray}
\end{widetext}

\begin{figure}[h]
\includegraphics[width=0.9\textwidth]{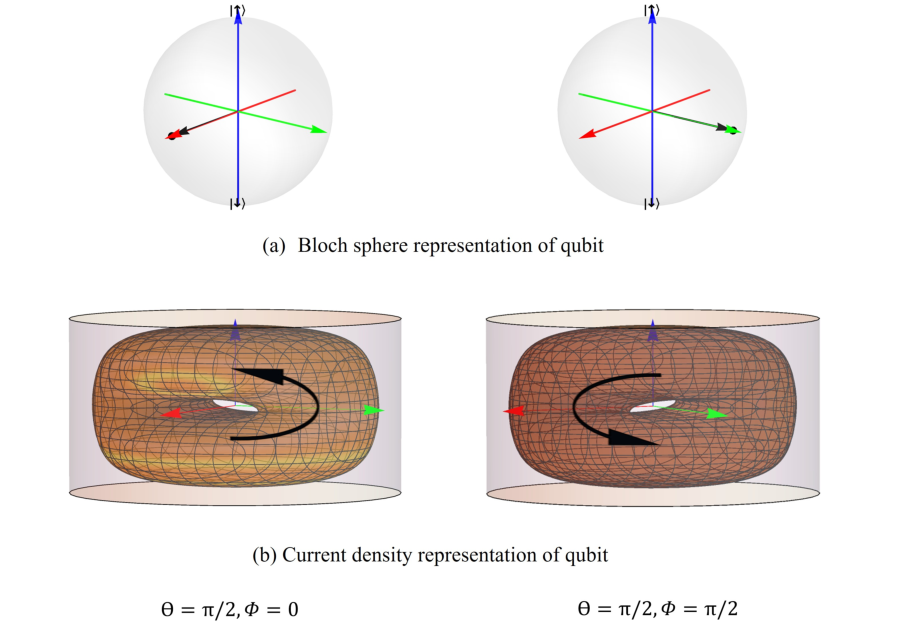}%
\caption{\label{fig:fig2}
(a) Bloch--sphere representation of a qubit as an abstract state vector.
(b) Three--dimensional contour plots of the current density for the wave--spin qubit, drawn at two--thirds of the peak value, reveal toroidal topologies circulating around $\phi=\Phi = 0$ (red, $x$--axis) and $\phi=\Phi = \tfrac{\pi}{2}$ (green, $y$--axis). These configurations highlight the phase--dependent character of the wave--spin qubit. Parameters: cylindrical cavity with $R = 8~\text{nm}$, $d = 4~\text{nm}$, $U = 10~\text{meV}$. The spin--up and spin--down components are degenerate in the ground state $(n l m = 101)$, with $\mathcal{E}_{101} - m_{e}c^2 = 8.06~\text{meV}$.}
\end{figure}

The transverse components $(j_\rho,j_\phi)$ vary as $(\sin(\Phi-\phi),-\cos(\Phi-\phi))$, i.e.\ along the unit tangent at azimuth $\phi=\Phi$. Thus, in the $xy$--plane the current points at angle $\Phi-\phi+\pi/2$, making the relative phase $\Phi$ directly set the local current orientation. Together with $\Theta$, this phase fixes the current configuration. Figure~\ref{fig:fig2}(b) illustrates this geometry: the current circulates around an axis aligned with $\phi=\Phi$. In the wave--entity picture, the superposition produces a coherent and well--defined current pattern, providing a physical representation of the qubit state traditionally depicted by vectors on the Bloch sphere in Fig.~\ref{fig:fig2}(a).

\section{\label{sec:InteractionWithField}Interaction with external fields}
The wave--entity framework establishes a distinct ontology for the electron relative to the standard wave--particle duality. This difference is not merely interpretive; it entails physical consequences accessible to experiment. In particular, interactions with structured electromagnetic fields and long--range couplings become sensitive to the local vector structure of the current~\cite{Taylor2003LongLived}.

To place the two views on equal footing, we compare their predictions for the magnetic moment. Within the wave--entity framework, the magnetic--moment density follows from classical electromagnetism as
\begin{equation}
\label{eq:Mdef}
\bm{M}(\bm{r}) = \tfrac{1}{2}\,\bm{r}\times\bm{j}(\bm{r}),
\end{equation}
which in cylindrical coordinates gives
\begin{equation}
\label{eq:Mcomponents}
\bm{M}(\bm{r}) = \tfrac{1}{2}\Big[-z\,j_{\phi}\,\hat{\bm{\rho}} + \big(z\,j_{\rho}-\rho\,j_{z}\big)\,\hat{\bm{\phi}} + \rho\,j_{\phi}\,\hat{\bm{z}}\Big].
\end{equation}
For the spin--up ground state, Eq.~\ref{current101} yields
\begin{equation}
\label{eq:M_we}
\bm{M}^{(\mathrm{wave\text{--}entity})}(\rho,\phi,z) =
\begin{cases}
\displaystyle \mu_{B}\, N^{2} \!\left[z\, \zeta J_{0}(\zeta \rho)\,J_{1}(\zeta \rho)\,\cos^{2}(kz)\,\hat{\bm{\rho}}
- \rho\,\zeta J_{0}(\zeta \rho)\,J_{1}(\zeta \rho)\,\cos^{2}(kz)\,\hat{\bm{z}} \right], & \text{Region I}, \\[2ex]
\displaystyle  \mu_{B} \, \kappa^{2} N^{2} \!\left[z\,\xi K_{0}(\xi \rho)\,K_{1}(\xi \rho)\,\cos^{2}(kz)\,\hat{\bm{\rho}}
- \rho\,\xi K_{0}(\xi \rho)\,K_{1}(\xi \rho)\,\cos^{2}(kz)\,\hat{\bm{z}} \right], & \text{Region II},
\end{cases}
\end{equation}
after substituting $\eta$ from Eq.~\ref{eta2} and introducing the Bohr magneton $\mu_{B} = e\hbar/(2m_{e})$. The resulting azimuthal current $j_{\phi}$ generates a poloidal magnetization.

By contrast, in the standard wave--particle treatment the electron carries an intrinsic magnetic dipole of magnitude $\mu \approx \mu_{B}$ aligned with the abstract spin axis, and the magnetization density is the dipole weighted by local probability density,
\begin{equation}
\label{eq:M_wp}
\bm{M}^{(\mathrm{wave\text{--}particle})}(\rho,\phi,z) =
-\,\mu_{B}\,\psi^\dagger(\rho,\phi,z)\,\psi(\rho,\phi,z)\,\hat{\bm{z}}
= \begin{cases}
\displaystyle -\,\mu_{B} N^{2} J_{0}(\zeta \rho)^{2}\cos^{2}(kz)\,\hat{\bm{z}}, & \text{Region I}, \\[2ex]
\displaystyle -\,\mu_{B} \kappa^{2} N^{2} K_{0}(\xi \rho)^{2}\cos^{2}(kz)\,\hat{\bm{z}}, & \text{Region II}.
\end{cases}
\end{equation}
This description is strictly axial: orientation is fixed along $-\hat{\bm{z}}$ and spatial variation enters only through the scalar probability amplitude. In the wave--entity framework, by contrast, circulating currents yield a poloidal magnetization with intertwined radial and axial components.

\begin{figure}[h]
  \centering
  \includegraphics[width=0.66\textwidth]{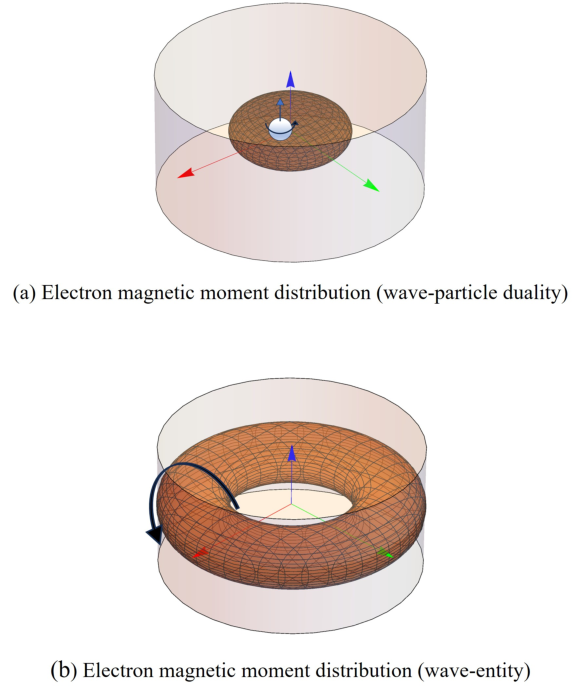}
\caption{\label{fig:MomentContours}
Contour plots of magnetic--moment density magnitudes for the ground state $(101)$ 
in a cylindrical cavity ($R=8~\mathrm{nm}$, $d=4~\mathrm{nm}$, $U=10~\mathrm{meV}$).
(a) Wave--particle picture: the moment density mirrors the probability distribution, 
resulting in a genus--0, sphere--like topology.  
(b) Wave--entity picture: the current density generates a genus--1 toroidal topology, 
with the corresponding poloidal vector pattern confined to the meridional plane, 
as described in Eq.~\ref{eq:M_we}.  
Contours are drawn at two--thirds of the maximum.
}
\end{figure}

For visualization we compare magnitudes,
\begin{equation}
\begin{aligned}
  \bigl|\bm{M}^{(\mathrm{wave\text{--}particle})}\bigr|
  &=
  \begin{cases}
  \displaystyle \mu_{B}\,N^{2}\,J_{0}(\zeta \rho)^{2}\cos^{2}(kz), & \text{Region I},\\[1ex]
  \displaystyle  \mu_{B}\,\kappa^{2} N^{2}\,K_{0}(\xi \rho)^{2}\cos^{2}(kz), & \text{Region II},
  \end{cases} \\[1ex]
  \bigl|\bm{M}^{(\mathrm{wave\text{--}entity})}\bigr|
  &=
  \begin{cases}
  \displaystyle  \mu_{B}\, N^{2} \zeta\,\sqrt{\rho^{2}+z^{2}}\;J_{0}(\zeta \rho)J_{1}(\zeta \rho)\cos^{2}(kz), & \text{Region I},\\[1ex]
  \displaystyle \mu_{B}\, \kappa^{2} N^{2} \xi\,\sqrt{\rho^{2}+z^{2}}\;K_{0}(\xi \rho)K_{1}(\xi \rho)\cos^{2}(kz), & \text{Region II}.
  \end{cases}
  \label{eq:Magnitudes}
\end{aligned}
\end{equation}
which reveals the characteristic contrast seen in the contour plots of Fig.~\ref{fig:MomentContours}. In the wave--particle picture the moment density mirrors the probability distribution, resulting in a genus--0, sphere--like topology. In contrast, the wave--entity expression reflects the current density, giving rise to a genus--1 toroidal topology. The accompanying poloidal vector pattern, confined to the meridional plane, 
is described explicitly in Eq.~\ref{eq:M_we}.

Despite local differences, integrating the moment density yields
\begin{equation}  \label{eq:mu_we}
  \bm{\mu}^{(\mathrm{wave\text{--}entity})}
  =\int_{0}^{\infty}\!\!\int_{0}^{2\pi}\!\!\int_{-d}^{d}
  \bm{M}^{(\mathrm{wave\text{--}entity})}(\rho,\phi,z)\,\rho\,d\rho\,d\phi\,dz
  =-\,\mu_{B}\,\hat{\bm{z}},
\end{equation}
with the $\hat{\bm{\rho}}$ contribution vanishing by odd parity in $z$, and radial integrals
\begin{equation} \label{eq:bessel_int}
  \int_{0}^{R}\! J_{0}(\zeta \rho)\,J_{1}(\zeta \rho)\,\rho^{2}\,d\rho
  =\frac{R^{2}}{2\zeta}\,J_{1}(\zeta R)^{2},\qquad
  \int_{R}^{\infty}\! K_{0}(\xi \rho)\,K_{1}(\xi \rho)\,\rho^{2}\,d\rho
  =\frac{R^{2}}{2\xi}\,K_{1}(\xi R)^{2}.
\end{equation}
Thus $|\bm{\mu}^{(\mathrm{wave\text{--}entity})}|=\mu_B$, recovering $g\approx 2$. In the wave--particle framework,
\begin{equation}
\bm{\mu}^{(\mathrm{wave\text{--}particle})}=-\,\mu_{B}\,\hat{\bm{z}}.
\end{equation}
Both approaches thus agree on the total moment in uniform fields, but differ in local vector structure; those differences are exposed by structured fields (gradients, curvature, or nontrivial vector potentials) and connect naturally to Aharonov--Bohm--like energy responses developed in our companion work.

\section{\label{sec:Conclusions}Conclusions}
Within the wave--entity framework, a spin qubit is not an abstract binary label but a tangible, spatially extended circulation of current density. This enhances the traditional Bloch--sphere representation with a coherent, physically real wave--spin configuration. The qubit phase $\Phi$ emerges as a genuine physical parameter: it directly orients this circulation and ties phase control to real--space current flow.

From these currents we obtain the magnetic--moment density, which in the wave--entity picture assumes a poloidal form with nodal structure---by contrast with the strictly axial form of the conventional wave--particle model. Both frameworks yield the same total magnetic moment (one Bohr magneton) and are therefore indistinguishable in uniform external fields. Yet the local topologies differ, leading to distinct couplings with structured external fields and with spin--spin interactions. This motivates targeted experiments---closely allied to Aharonov--Bohm--like protocols---to discriminate between the frameworks.

More broadly, the wave--entity framework offers a physically grounded and deterministic ontology that may help clarify the connections between quantum and classical descriptions of matter, while providing concrete, testable predictions for current--field interactions in engineered structures.

\section*{\label{sec:Acknowledgements}Acknowledgments}
We gratefully acknowledge the reviewers’ insightful comments and constructive feedback, which have substantially improved the clarity and rigor of this work. We also thank the editorial team for their thoughtful guidance throughout the review process.

\bibliography{Spin} 

\providecommand{\noopsort}[1]{}\providecommand{\singleletter}[1]{#1}%
\begin{thebibliography}{33}%
\makeatletter
\providecommand \@ifxundefined [1]{%
 \@ifx{#1\undefined}
}%
\providecommand \@ifnum [1]{%
 \ifnum #1\expandafter \@firstoftwo
 \else \expandafter \@secondoftwo
 \fi
}%
\providecommand \@ifx [1]{%
 \ifx #1\expandafter \@firstoftwo
 \else \expandafter \@secondoftwo
 \fi
}%
\providecommand \natexlab [1]{#1}%
\providecommand \enquote  [1]{``#1''}%
\providecommand \bibnamefont  [1]{#1}%
\providecommand \bibfnamefont [1]{#1}%
\providecommand \citenamefont [1]{#1}%
\providecommand \href@noop [0]{\@secondoftwo}%
\providecommand \href [0]{\begingroup \@sanitize@url \@href}%
\providecommand \@href[1]{\@@startlink{#1}\@@href}%
\providecommand \@@href[1]{\endgroup#1\@@endlink}%
\providecommand \@sanitize@url [0]{\catcode `\\12\catcode `\$12\catcode
  `\&12\catcode `\#12\catcode `\^12\catcode `\_12\catcode `\%12\relax}%
\providecommand \@@startlink[1]{}%
\providecommand \@@endlink[0]{}%
\providecommand \url  [0]{\begingroup\@sanitize@url \@url }%
\providecommand \@url [1]{\endgroup\@href {#1}{\urlprefix }}%
\providecommand \urlprefix  [0]{URL }%
\providecommand \Eprint [0]{\href }%
\providecommand \doibase [0]{https://doi.org/}%
\providecommand \selectlanguage [0]{\@gobble}%
\providecommand \bibinfo  [0]{\@secondoftwo}%
\providecommand \bibfield  [0]{\@secondoftwo}%
\providecommand \translation [1]{[#1]}%
\providecommand \BibitemOpen [0]{}%
\providecommand \bibitemStop [0]{}%
\providecommand \bibitemNoStop [0]{.\EOS\space}%
\providecommand \EOS [0]{\spacefactor3000\relax}%
\providecommand \BibitemShut  [1]{\csname bibitem#1\endcsname}%
\let\auto@bib@innerbib\@empty
\bibitem [{\citenamefont {Gao}(2022)}]{GaoJOPCO22}%
  \BibitemOpen
  \bibfield  {author} {\bibinfo {author} {\bibfnamefont {J.}~\bibnamefont
  {Gao}},\ }\href@noop {} {\bibfield  {journal} {\bibinfo  {journal} {J. Phys.
  Commun.}\ }\textbf {\bibinfo {volume} {6}},\ \bibinfo {pages} {081001}
  (\bibinfo {year} {2022})}\BibitemShut {NoStop}%
\bibitem [{\citenamefont {Gao}\ and\ \citenamefont
  {Shen}(2024)}]{EntropyEvaSpin2024}%
  \BibitemOpen
  \bibfield  {author} {\bibinfo {author} {\bibfnamefont {J.}~\bibnamefont
  {Gao}}\ and\ \bibinfo {author} {\bibfnamefont {F.}~\bibnamefont {Shen}},\
  }\bibfield  {title} {\bibinfo {title} {Evanescent electron wave-spin},\
  }\bibfield  {journal} {\bibinfo  {journal} {Entropy}\ }\textbf {\bibinfo
  {volume} {26}},\ \href {https://doi.org/10.3390/e26090789}
  {10.3390/e26090789} (\bibinfo {year} {2024})\BibitemShut {NoStop}%
\bibitem [{\citenamefont {Jackson}(1999)}]{JDJackson1999}%
  \BibitemOpen
  \bibfield  {author} {\bibinfo {author} {\bibfnamefont {J.~D.}\ \bibnamefont
  {Jackson}},\ }\href {https://search.library.wisc.edu/catalog/999849741702121}
  {\emph {\bibinfo {title} {Classical Electrodynamics}}}\ (\bibinfo
  {publisher} {Third edition. New York : Wiley, 1999},\ \bibinfo {year}
  {1999})\BibitemShut {NoStop}%
\bibitem [{\citenamefont
  {Jönsson}(1961)}]{jonsson1961elektroneninterferenzen}%
  \BibitemOpen
  \bibfield  {author} {\bibinfo {author} {\bibfnamefont {C.}~\bibnamefont
  {Jönsson}},\ }\bibfield  {title} {\bibinfo {title} {Elektroneninterferenzen
  an mehreren künstlich hergestellten feinspalten},\ }\href
  {https://doi.org/10.1007/BF01342460} {\bibfield  {journal} {\bibinfo
  {journal} {Zeitschrift für Physik}\ }\textbf {\bibinfo {volume} {161}},\
  \bibinfo {pages} {454} (\bibinfo {year} {1961})}\BibitemShut {NoStop}%
\bibitem [{\citenamefont {Jönsson}(1974)}]{jonsson1961electron}%
  \BibitemOpen
  \bibfield  {author} {\bibinfo {author} {\bibfnamefont {C.}~\bibnamefont
  {Jönsson}},\ }\bibfield  {title} {\bibinfo {title} {Electron diffraction at
  multiple slits},\ }\href {https://doi.org/10.1119/1.1987592} {\bibfield
  {journal} {\bibinfo  {journal} {American Journal of Physics}\ }\textbf
  {\bibinfo {volume} {42}},\ \bibinfo {pages} {4} (\bibinfo {year} {1974})},\
  \bibinfo {note} {english translation of the 1961 German paper}\BibitemShut
  {NoStop}%
\bibitem [{\citenamefont {Tonomura}\ \emph {et~al.}(1989)\citenamefont
  {Tonomura}, \citenamefont {Endo}, \citenamefont {Matsuda}, \citenamefont
  {Kawasaki},\ and\ \citenamefont {Ezawa}}]{tonomura1989demonstration}%
  \BibitemOpen
  \bibfield  {author} {\bibinfo {author} {\bibfnamefont {A.}~\bibnamefont
  {Tonomura}}, \bibinfo {author} {\bibfnamefont {J.}~\bibnamefont {Endo}},
  \bibinfo {author} {\bibfnamefont {T.}~\bibnamefont {Matsuda}}, \bibinfo
  {author} {\bibfnamefont {T.}~\bibnamefont {Kawasaki}},\ and\ \bibinfo
  {author} {\bibfnamefont {H.}~\bibnamefont {Ezawa}},\ }\bibfield  {title}
  {\bibinfo {title} {Demonstration of single-electron buildup of an
  interference pattern},\ }\href {https://doi.org/10.1119/1.16104} {\bibfield
  {journal} {\bibinfo  {journal} {American Journal of Physics}\ }\textbf
  {\bibinfo {volume} {57}},\ \bibinfo {pages} {117} (\bibinfo {year}
  {1989})}\BibitemShut {NoStop}%
\bibitem [{\citenamefont {Davisson}\ and\ \citenamefont
  {Germer}(1927)}]{davisson1927diffraction}%
  \BibitemOpen
  \bibfield  {author} {\bibinfo {author} {\bibfnamefont {C.}~\bibnamefont
  {Davisson}}\ and\ \bibinfo {author} {\bibfnamefont {L.~H.}\ \bibnamefont
  {Germer}},\ }\bibfield  {title} {\bibinfo {title} {Diffraction of electrons
  by a crystal of nickel},\ }\href {https://doi.org/10.1103/PhysRev.30.705}
  {\bibfield  {journal} {\bibinfo  {journal} {Physical Review}\ }\textbf
  {\bibinfo {volume} {30}},\ \bibinfo {pages} {705} (\bibinfo {year}
  {1927})}\BibitemShut {NoStop}%
\bibitem [{\citenamefont {Thomson}\ and\ \citenamefont
  {Reid}(1927)}]{thomson1927diffraction}%
  \BibitemOpen
  \bibfield  {author} {\bibinfo {author} {\bibfnamefont {G.~P.}\ \bibnamefont
  {Thomson}}\ and\ \bibinfo {author} {\bibfnamefont {A.}~\bibnamefont {Reid}},\
  }\bibfield  {title} {\bibinfo {title} {Diffraction of cathode rays by a thin
  film},\ }\href {https://doi.org/10.1038/119890b0} {\bibfield  {journal}
  {\bibinfo  {journal} {Nature}\ }\textbf {\bibinfo {volume} {119}},\ \bibinfo
  {pages} {890} (\bibinfo {year} {1927})}\BibitemShut {NoStop}%
\bibitem [{\citenamefont {Huang}\ \emph {et~al.}(2024)\citenamefont {Huang}
  \emph {et~al.}}]{Huang2024Nature}%
  \BibitemOpen
  \bibfield  {author} {\bibinfo {author} {\bibfnamefont {J.~Y.}\ \bibnamefont
  {Huang}} \emph {et~al.},\ }\bibfield  {title} {\bibinfo {title}
  {High-fidelity spin qubit operation and algorithmic performance above 1
  {K}},\ }\bibfield  {journal} {\bibinfo  {journal} {Nature}\ }\textbf
  {\bibinfo {volume} {629}},\ \href
  {https://doi.org/10.1038/s41586-024-07160-2} {10.1038/s41586-024-07160-2}
  (\bibinfo {year} {2024})\BibitemShut {NoStop}%
\bibitem [{\citenamefont {Hu}\ \emph {et~al.}(2025)\citenamefont {Hu} \emph
  {et~al.}}]{Hu2025iComputing}%
  \BibitemOpen
  \bibfield  {author} {\bibinfo {author} {\bibfnamefont {G.}~\bibnamefont {Hu}}
  \emph {et~al.},\ }\bibfield  {title} {\bibinfo {title} {Single-electron spin
  qubits in silicon for quantum information processing: From fundamentals to
  large-scale integration},\ }\href {https://doi.org/10.34133/icomputing.0115}
  {\bibfield  {journal} {\bibinfo  {journal} {iComputing}\ }\textbf {\bibinfo
  {volume} {X}},\ \bibinfo {pages} {0115} (\bibinfo {year} {2025})}\BibitemShut
  {NoStop}%
\bibitem [{\citenamefont {Meunier}\ \emph {et~al.}(2025)\citenamefont {Meunier}
  \emph {et~al.}}]{Meunier2025EPJA}%
  \BibitemOpen
  \bibfield  {author} {\bibinfo {author} {\bibfnamefont {T.}~\bibnamefont
  {Meunier}} \emph {et~al.},\ }\bibfield  {title} {\bibinfo {title} {Silicon
  spin qubits: a viable path towards industrial quantum computers},\ }\bibfield
   {journal} {\bibinfo  {journal} {The European Physical Journal A}\ }\href
  {https://doi.org/10.1140/epja/s10050-025-01514-8}
  {10.1140/epja/s10050-025-01514-8} (\bibinfo {year} {2025})\BibitemShut
  {NoStop}%
\bibitem [{\citenamefont {Cai}\ \emph {et~al.}(2025)\citenamefont {Cai} \emph
  {et~al.}}]{Cai2025PRApplied}%
  \BibitemOpen
  \bibfield  {author} {\bibinfo {author} {\bibfnamefont {R.}~\bibnamefont
  {Cai}} \emph {et~al.},\ }\bibfield  {title} {\bibinfo {title} {Ultrafast
  switchable spin--orbit coupling for silicon spin qubits using tunable
  magnetic textures},\ }\href
  {https://doi.org/10.1103/PhysRevApplied.23.024048} {\bibfield  {journal}
  {\bibinfo  {journal} {Physical Review Applied}\ }\textbf {\bibinfo {volume}
  {23}},\ \bibinfo {pages} {024048} (\bibinfo {year} {2025})},\ \bibinfo {note}
  {see also arXiv:2310.17993}\BibitemShut {NoStop}%
\bibitem [{\citenamefont {Dirac}(1928)}]{Dirac28}%
  \BibitemOpen
  \bibfield  {author} {\bibinfo {author} {\bibfnamefont {P.~A.~M.}\
  \bibnamefont {Dirac}},\ }\href@noop {} {\bibfield  {journal} {\bibinfo
  {journal} {Proceedings of the Royal Society A: Mathematical,Physical and
  Engineering Sciences}\ }\textbf {\bibinfo {volume} {117}},\ \bibinfo {pages}
  {610} (\bibinfo {year} {1928})}\BibitemShut {NoStop}%
\bibitem [{\citenamefont {Dirac}(1930)}]{dirac1930principles}%
  \BibitemOpen
  \bibfield  {author} {\bibinfo {author} {\bibfnamefont {P.~A.~M.}\
  \bibnamefont {Dirac}},\ }\href@noop {} {\emph {\bibinfo {title} {The
  Principles of Quantum Mechanics}}}\ (\bibinfo  {publisher} {Oxford University
  Press},\ \bibinfo {year} {1930})\BibitemShut {NoStop}%
\bibitem [{\citenamefont {Ohanian}(1986)}]{Ohanian86}%
  \BibitemOpen
  \bibfield  {author} {\bibinfo {author} {\bibfnamefont {H.~C.}\ \bibnamefont
  {Ohanian}},\ }\href@noop {} {\bibfield  {journal} {\bibinfo  {journal} {Am.
  J. Phys.}\ }\textbf {\bibinfo {volume} {54}},\ \bibinfo {pages} {6} (\bibinfo
  {year} {1986})}\BibitemShut {NoStop}%
\bibitem [{\citenamefont {Paiva}\ \emph {et~al.}(2023)\citenamefont {Paiva},
  \citenamefont {Aharonov}, \citenamefont {Tollaksen},\ and\ \citenamefont
  {Waegell}}]{Paiva2023NJP}%
  \BibitemOpen
  \bibfield  {author} {\bibinfo {author} {\bibfnamefont {I.~L.}\ \bibnamefont
  {Paiva}}, \bibinfo {author} {\bibfnamefont {Y.}~\bibnamefont {Aharonov}},
  \bibinfo {author} {\bibfnamefont {J.}~\bibnamefont {Tollaksen}},\ and\
  \bibinfo {author} {\bibfnamefont {M.}~\bibnamefont {Waegell}},\ }\bibfield
  {title} {\bibinfo {title} {Aharonov--bohm effect with an effective
  complex-valued vector potential},\ }\href
  {https://doi.org/10.1088/1367-2630/acd4dd} {\bibfield  {journal} {\bibinfo
  {journal} {New Journal of Physics}\ }\textbf {\bibinfo {volume} {25}},\
  \bibinfo {pages} {053017} (\bibinfo {year} {2023})}\BibitemShut {NoStop}%
\bibitem [{\citenamefont {Chen}\ \emph {et~al.}(2023)\citenamefont {Chen},
  \citenamefont {Zhu},\ and\ \citenamefont {Chen}}]{Chen2023SpinAB}%
  \BibitemOpen
  \bibfield  {author} {\bibinfo {author} {\bibfnamefont {Y.}~\bibnamefont
  {Chen}}, \bibinfo {author} {\bibfnamefont {W.}~\bibnamefont {Zhu}},\ and\
  \bibinfo {author} {\bibfnamefont {Z.-D.}\ \bibnamefont {Chen}},\ }\bibfield
  {title} {\bibinfo {title} {Spin vector potential and spin aharonov--bohm
  effect},\ }\href {https://arxiv.org/abs/2211.07178} {\bibfield  {journal}
  {\bibinfo  {journal} {arXiv preprint}\ } (\bibinfo {year} {2023})},\ \Eprint
  {https://arxiv.org/abs/2211.07178} {arXiv:2211.07178 [quant-ph]} \BibitemShut
  {NoStop}%
\bibitem [{\citenamefont {Loss}\ and\ \citenamefont
  {DiVincenzo}(1998)}]{loss1998quantum}%
  \BibitemOpen
  \bibfield  {author} {\bibinfo {author} {\bibfnamefont {D.}~\bibnamefont
  {Loss}}\ and\ \bibinfo {author} {\bibfnamefont {D.~P.}\ \bibnamefont
  {DiVincenzo}},\ }\bibfield  {title} {\bibinfo {title} {Quantum computation
  with quantum dots},\ }\href@noop {} {\bibfield  {journal} {\bibinfo
  {journal} {Physical Review A}\ }\textbf {\bibinfo {volume} {57}},\ \bibinfo
  {pages} {120} (\bibinfo {year} {1998})}\BibitemShut {NoStop}%
\bibitem [{\citenamefont {Hanson}\ \emph {et~al.}(2007)\citenamefont {Hanson},
  \citenamefont {Kouwenhoven}, \citenamefont {Petta}, \citenamefont {Tarucha},\
  and\ \citenamefont {Vandersypen}}]{hanson2007spins}%
  \BibitemOpen
  \bibfield  {author} {\bibinfo {author} {\bibfnamefont {R.}~\bibnamefont
  {Hanson}}, \bibinfo {author} {\bibfnamefont {L.~P.}\ \bibnamefont
  {Kouwenhoven}}, \bibinfo {author} {\bibfnamefont {J.~R.}\ \bibnamefont
  {Petta}}, \bibinfo {author} {\bibfnamefont {S.}~\bibnamefont {Tarucha}},\
  and\ \bibinfo {author} {\bibfnamefont {L.~M.~K.}\ \bibnamefont
  {Vandersypen}},\ }\bibfield  {title} {\bibinfo {title} {Spins in few-electron
  quantum dots},\ }\href@noop {} {\bibfield  {journal} {\bibinfo  {journal}
  {Reviews of Modern Physics}\ }\textbf {\bibinfo {volume} {79}},\ \bibinfo
  {pages} {1217} (\bibinfo {year} {2007})}\BibitemShut {NoStop}%
\bibitem [{\citenamefont {Lobanova}\ and\ \citenamefont
  {Ivanov}(2004)}]{lobanova2004cylindrical}%
  \BibitemOpen
  \bibfield  {author} {\bibinfo {author} {\bibfnamefont {O.~R.}\ \bibnamefont
  {Lobanova}}\ and\ \bibinfo {author} {\bibfnamefont {A.~I.}\ \bibnamefont
  {Ivanov}},\ }\href@noop {} {} (\bibinfo {year} {2004}),\ \Eprint
  {https://arxiv.org/abs/quant-ph/0411150} {arXiv:quant-ph/0411150 [quant-ph]}
  \BibitemShut {NoStop}%
\bibitem [{\citenamefont {Lambert}\ \emph {et~al.}(2013)\citenamefont
  {Lambert}, \citenamefont {Chen}, \citenamefont {Cheng}, \citenamefont {Li},
  \citenamefont {Chen},\ and\ \citenamefont
  {Nori}}]{Lambert2013QuantumBiology}%
  \BibitemOpen
  \bibfield  {author} {\bibinfo {author} {\bibfnamefont {N.}~\bibnamefont
  {Lambert}}, \bibinfo {author} {\bibfnamefont {Y.-N.}\ \bibnamefont {Chen}},
  \bibinfo {author} {\bibfnamefont {Y.-C.}\ \bibnamefont {Cheng}}, \bibinfo
  {author} {\bibfnamefont {C.-M.}\ \bibnamefont {Li}}, \bibinfo {author}
  {\bibfnamefont {G.-Y.}\ \bibnamefont {Chen}},\ and\ \bibinfo {author}
  {\bibfnamefont {F.}~\bibnamefont {Nori}},\ }\bibfield  {title} {\bibinfo
  {title} {Quantum biology},\ }\href {https://doi.org/10.1038/nphys2474}
  {\bibfield  {journal} {\bibinfo  {journal} {Nature Physics}\ }\textbf
  {\bibinfo {volume} {9}},\ \bibinfo {pages} {10} (\bibinfo {year}
  {2013})}\BibitemShut {NoStop}%
\bibitem [{\citenamefont {Onizhuk}\ \emph {et~al.}(2025)\citenamefont {Onizhuk}
  \emph {et~al.}}]{Onizhuk2025RMP}%
  \BibitemOpen
  \bibfield  {author} {\bibinfo {author} {\bibfnamefont {M.}~\bibnamefont
  {Onizhuk}} \emph {et~al.},\ }\bibfield  {title} {\bibinfo {title}
  {Colloquium: Decoherence of solid-state spin qubits},\ }\href
  {https://doi.org/10.1103/RevModPhys.97.021001} {\bibfield  {journal}
  {\bibinfo  {journal} {Reviews of Modern Physics}\ }\textbf {\bibinfo {volume}
  {97}},\ \bibinfo {pages} {021001} (\bibinfo {year} {2025})}\BibitemShut
  {NoStop}%
\bibitem [{\citenamefont {Bloch}(1946)}]{bloch1946nuclear}%
  \BibitemOpen
  \bibfield  {author} {\bibinfo {author} {\bibfnamefont {F.}~\bibnamefont
  {Bloch}},\ }\bibfield  {title} {\bibinfo {title} {Nuclear induction},\ }\href
  {https://doi.org/10.1103/PhysRev.70.460} {\bibfield  {journal} {\bibinfo
  {journal} {Physical Review}\ }\textbf {\bibinfo {volume} {70}},\ \bibinfo
  {pages} {460} (\bibinfo {year} {1946})}\BibitemShut {NoStop}%
\bibitem [{\citenamefont {Nielsen}\ and\ \citenamefont
  {Chuang}(2010)}]{nielsen2010quantum}%
  \BibitemOpen
  \bibfield  {author} {\bibinfo {author} {\bibfnamefont {M.~A.}\ \bibnamefont
  {Nielsen}}\ and\ \bibinfo {author} {\bibfnamefont {I.~L.}\ \bibnamefont
  {Chuang}},\ }\href@noop {} {\emph {\bibinfo {title} {Quantum Computation and
  Quantum Information}}},\ \bibinfo {edition} {10th}\ ed.\ (\bibinfo
  {publisher} {Cambridge University Press},\ \bibinfo {year}
  {2010})\BibitemShut {NoStop}%
\bibitem [{\citenamefont {Song}\ \emph {et~al.}(2024)\citenamefont {Song},
  \citenamefont {Yun}, \citenamefont {Kim},\ and\ \citenamefont
  {et~al.}}]{Song2024npjQI}%
  \BibitemOpen
  \bibfield  {author} {\bibinfo {author} {\bibfnamefont {Y.}~\bibnamefont
  {Song}}, \bibinfo {author} {\bibfnamefont {J.}~\bibnamefont {Yun}}, \bibinfo
  {author} {\bibfnamefont {J.}~\bibnamefont {Kim}},\ and\ \bibinfo {author}
  {\bibnamefont {et~al.}},\ }\bibfield  {title} {\bibinfo {title} {Coherence of
  a field gradient driven singlet--triplet qubit coupled to multielectron spin
  states in $^{28}$si/sige},\ }\bibfield  {journal} {\bibinfo  {journal} {npj
  Quantum Information}\ }\textbf {\bibinfo {volume} {10}},\ \href
  {https://doi.org/10.1038/s41534-024-00869-y} {10.1038/s41534-024-00869-y}
  (\bibinfo {year} {2024})\BibitemShut {NoStop}%
\bibitem [{\citenamefont {Unseld}\ \emph {et~al.}(2025)\citenamefont {Unseld}
  \emph {et~al.}}]{Unseld2025NatComm}%
  \BibitemOpen
  \bibfield  {author} {\bibinfo {author} {\bibfnamefont {L.}~\bibnamefont
  {Unseld}} \emph {et~al.},\ }\bibfield  {title} {\bibinfo {title} {Baseband
  control of spin qubits},\ }\bibfield  {journal} {\bibinfo  {journal} {Nature
  Communications}\ }\textbf {\bibinfo {volume} {16}},\ \href
  {https://doi.org/10.1038/s41467-025-54251-9} {10.1038/s41467-025-54251-9}
  (\bibinfo {year} {2025})\BibitemShut {NoStop}%
\bibitem [{\citenamefont {Cleve}\ \emph {et~al.}(1998)\citenamefont {Cleve},
  \citenamefont {Ekert}, \citenamefont {Macchiavello},\ and\ \citenamefont
  {Mosca}}]{cleve1998quantum}%
  \BibitemOpen
  \bibfield  {author} {\bibinfo {author} {\bibfnamefont {R.}~\bibnamefont
  {Cleve}}, \bibinfo {author} {\bibfnamefont {A.}~\bibnamefont {Ekert}},
  \bibinfo {author} {\bibfnamefont {C.}~\bibnamefont {Macchiavello}},\ and\
  \bibinfo {author} {\bibfnamefont {M.}~\bibnamefont {Mosca}},\ }\bibfield
  {title} {\bibinfo {title} {Quantum algorithms revisited},\ }\href
  {https://doi.org/10.1098/rspa.1998.0164} {\bibfield  {journal} {\bibinfo
  {journal} {Proceedings of the Royal Society of London. Series A:
  Mathematical, Physical and Engineering Sciences}\ }\textbf {\bibinfo {volume}
  {454}},\ \bibinfo {pages} {339} (\bibinfo {year} {1998})}\BibitemShut
  {NoStop}%
\bibitem [{\citenamefont {Coppersmith}(2002)}]{coppersmith2002approximate}%
  \BibitemOpen
  \bibfield  {author} {\bibinfo {author} {\bibfnamefont {D.}~\bibnamefont
  {Coppersmith}},\ }\bibfield  {title} {\bibinfo {title} {An approximate
  fourier transform useful in quantum factoring},\ }\href
  {https://arxiv.org/abs/quant-ph/0201067} {\bibfield  {journal} {\bibinfo
  {journal} {arXiv preprint quant-ph/0201067}\ } (\bibinfo {year}
  {2002})}\BibitemShut {NoStop}%
\bibitem [{\citenamefont {Shor}(1999)}]{shor1999polynomial}%
  \BibitemOpen
  \bibfield  {author} {\bibinfo {author} {\bibfnamefont {P.~W.}\ \bibnamefont
  {Shor}},\ }\bibfield  {title} {\bibinfo {title} {Polynomial-time algorithms
  for prime factorization and discrete logarithms on a quantum computer},\
  }\href {https://doi.org/10.1137/S0036144598347011} {\bibfield  {journal}
  {\bibinfo  {journal} {SIAM review}\ }\textbf {\bibinfo {volume} {41}},\
  \bibinfo {pages} {303} (\bibinfo {year} {1999})}\BibitemShut {NoStop}%
\bibitem [{\citenamefont {Grover}(1997)}]{grover1997quantum}%
  \BibitemOpen
  \bibfield  {author} {\bibinfo {author} {\bibfnamefont {L.~K.}\ \bibnamefont
  {Grover}},\ }\bibfield  {title} {\bibinfo {title} {Quantum mechanics helps in
  searching for a needle in a haystack},\ }\href
  {https://doi.org/10.1103/PhysRevLett.79.325} {\bibfield  {journal} {\bibinfo
  {journal} {Physical review letters}\ }\textbf {\bibinfo {volume} {79}},\
  \bibinfo {pages} {325} (\bibinfo {year} {1997})}\BibitemShut {NoStop}%
\bibitem [{\citenamefont {Nielsen}\ and\ \citenamefont
  {Chuang}(2002)}]{nielsen2002quantum}%
  \BibitemOpen
  \bibfield  {author} {\bibinfo {author} {\bibfnamefont {M.~A.}\ \bibnamefont
  {Nielsen}}\ and\ \bibinfo {author} {\bibfnamefont {I.~L.}\ \bibnamefont
  {Chuang}},\ }\bibfield  {title} {\bibinfo {title} {Quantum computation and
  quantum information},\ }\href {https://doi.org/10.1119/1.1463744} {\bibfield
  {journal} {\bibinfo  {journal} {American Journal of Physics}\ }\textbf
  {\bibinfo {volume} {70}},\ \bibinfo {pages} {558} (\bibinfo {year} {2002})},\
  \bibinfo {note} {book review}\BibitemShut {NoStop}%
\bibitem [{\citenamefont {Deutsch}\ and\ \citenamefont
  {Jozsa}(1992)}]{deutsch1992rapid}%
  \BibitemOpen
  \bibfield  {author} {\bibinfo {author} {\bibfnamefont {D.}~\bibnamefont
  {Deutsch}}\ and\ \bibinfo {author} {\bibfnamefont {R.}~\bibnamefont
  {Jozsa}},\ }\bibfield  {title} {\bibinfo {title} {Rapid solution of problems
  by quantum computation},\ }\href {https://doi.org/10.1098/rspa.1992.0167}
  {\bibfield  {journal} {\bibinfo  {journal} {Proceedings of the Royal Society
  of London. Series A: Mathematical and Physical Sciences}\ }\textbf {\bibinfo
  {volume} {439}},\ \bibinfo {pages} {553} (\bibinfo {year}
  {1992})}\BibitemShut {NoStop}%
\bibitem [{\citenamefont {Taylor}\ \emph {et~al.}(2003)\citenamefont {Taylor},
  \citenamefont {Marcus},\ and\ \citenamefont {Lukin}}]{Taylor2003LongLived}%
  \BibitemOpen
  \bibfield  {author} {\bibinfo {author} {\bibfnamefont {J.~M.}\ \bibnamefont
  {Taylor}}, \bibinfo {author} {\bibfnamefont {C.~M.}\ \bibnamefont {Marcus}},\
  and\ \bibinfo {author} {\bibfnamefont {M.~D.}\ \bibnamefont {Lukin}},\
  }\bibfield  {title} {\bibinfo {title} {Long-lived memory for mesoscopic
  quantum bits},\ }\href {https://doi.org/10.1103/PhysRevLett.90.206803}
  {\bibfield  {journal} {\bibinfo  {journal} {Physical Review Letters}\
  }\textbf {\bibinfo {volume} {90}},\ \bibinfo {pages} {206803} (\bibinfo
  {year} {2003})}\BibitemShut {NoStop}%
\end{thebibliography}%

\end{document}